\pdfoutput=1

\documentclass[pre,twocolumn,showpacs,aps,floats,floatfix]{revtex4-1}

\usepackage{natbib}
\usepackage[pdftex]{graphicx}
\usepackage{amssymb,amsmath}
\usepackage{amsfonts}
\usepackage{mathrsfs}
\usepackage[english]{babel}
\usepackage{epsfig}
\usepackage{epstopdf}
\usepackage{babelbib}
\usepackage{hyperref}
\usepackage{color}
\usepackage{tcolorbox}
\usepackage{tabularx}
\usepackage{array}
\usepackage{colortbl}
\tcbuselibrary{skins}
\usepackage{comment}
\usepackage{array}
\usepackage{bm}
\usepackage{mathtools}

\makeatletter
\newcommand*{\textoverline}[1]{$\overline{\hbox{#1}}\m@th$}
\makeatother

\begin{document}

\title{Spherical structure factor and classification of hyperuniformity on the sphere}
\author{An\v ze Lo\v sdorfer Bo\v zi\v c}
\affiliation{Department of Theoretical Physics, Jo\v{z}ef Stefan Institute, SI-1000 Ljubljana, Slovenia}
\email{anze.bozic@ijs.si}
\author{Simon \v Copar}
\affiliation{Faculty of Mathematics and Physics, University of Ljubljana, SI-1000 Ljubljana, Slovenia}

\date{\today}

\begin{abstract}
Understanding how particles are arranged on the surface of a sphere is not only central to numerous physical, biological, soft matter, and materials systems but also finds applications in computational problems, approximation theory, and analysis of geophysical and meteorological measurements. Objects that lie on a sphere experience constraints which are not present in Euclidean (flat) space and which influence both how the particles can be arranged as well as their statistical properties. These constraints, coupled with the curved geometry, require a careful extension of quantities used for the analysis of particle distributions in Euclidean space to distributions confined to the surface of a sphere. Here, we introduce a framework designed to analyze and classify structural order and disorder in particle distributions constrained to the sphere. The classification is based on the concept of hyperuniformity, which was first introduced $15$ years ago and since then studied extensively in Euclidean space, yet has only very recently been considered also for spherical surfaces. We employ a generalization of the structure factor on the sphere, related to the power spectrum of the corresponding multipole expansion of particle density distribution. The spherical structure factor is then shown to couple with cap number variance, a measure of density variations at different scales, allowing us to analytically derive different forms of the variance pertaining to different types of distributions. Based on these forms, we construct a classification of hyperuniformity for scale-free particle distributions on the sphere and show how it can be extended to include other distribution types as well. We demonstrate that hyperuniformity on the sphere can be defined either through a vanishing spherical structure factor at low multipole numbers or through a scaling of the cap number variance---in both cases extending the Euclidean definition, while at the same time pointing out crucial differences. Our work thus provides a comprehensive tool for detecting global, long-range order on spheres and for the analysis of spherical computational meshes, biological and synthetic spherical assemblies, and ordering phase transitions in spherically-distributed particles.
\end{abstract}

\maketitle

\section{Introduction}

Distributing many points on the surface of a two-dimensional sphere is a problem that has arisen in numerous guises, with the answers being far from trivial and depending strongly on the sought-for properties. Widely studied models on the sphere, such as the Tammes problem of the packing of hard circles and the Thomson problem of the arrangements of repulsive charged particles, have exact solutions only for select, small numbers of interacting particles~\cite{Saff1997,Pfender2004,Wales2006}. These models have also been extended to include, for instance, long-range power-law interactions~\cite{Bowick2006}, line-connected charges~\cite{Slosar2006}, or systems of particles interacting through soft potentials~\cite{Franzini2018,Miller2011}, with each extension introducing novel behaviour into the system. The question of arranging points on the sphere is also relevant to computational problems, numerical analysis, and approximation theory, e.g. to design meshes with desired characteristics~\cite{Brauchart2015,Hardin2016}. While interesting from a purely mathematical perspective, the study of distributions of particles on surfaces of non-trivial topology has found application in a broad range of physical, biological, and chemical systems, including arrangements of colloidal particles in colloidosomes and of protein subunits in capsids of spherical viruses, micro-patterning of spherical particles for use in photonic crystals, and structures of multielectron bubbles and lipid vesicles, to name just a few~\cite{Thompson2015,Bowick2009,ALB2013a,Roshal2015}.

Confining particles to the surface of a sphere introduces several constraints to their placement that are absent in Euclidean (flat) space. Both the curvature and the topology of the sphere play a role in the way particles can be arranged on it---most prominently, no regular lattice can be fit onto its surface, as the topology of the sphere requires $12$ pentagonal disclinations to be present~\cite{Bowick2009}. The compactness of the sphere also requires a careful treatment: instead of the straightforward thermodynamic limit,  configurations of finite (small) numbers of particles become more relevant, and both the number of particles and the size of the sphere need to be considered as two independent parameters~\cite{Hill1994,Post1986}. Typically, distributions of particles on the sphere are characterized by their equidistribution, quasi-uniformity, potential energy, and pair correlation function~\cite{Hardin2016,Giarritta1992,Giarritta1993,Fantoni2012}. Other measures of order and symmetry that are used in Euclidean space---and are often dependent on its infinite extent---are more difficult to immediately generalize to sets of particles on the surface of the sphere. Two such measures are in particular oft-used to characterize the behaviour of systems of particles and the order present in them: structure factor and, related to it, hyperuniformity.

Hyperuniform states of matter in Euclidean space of arbitrary dimension are correlated many-particle systems in which density variations (spatial fluctuations) are completely suppressed at very large length scales~\cite{Torquato2003,Torquato2018}. This in turn implies that the structure factor vanishes in the limit of small wave vectors (long wavelengths), in contrast to ordinary disordered systems. Equivalently, a hyperuniform system is one in which the variance of the number of particles contained within a spherical observation window grows more slowly than the window volume in the limit of large window sizes, and the system is spatially correlated at large distances. These two properties hold true not only for perfect crystals and quasicrystals~\cite{Torquato2003,Oguz2017} but also for certain states of disordered matter, such as, for instance, disordered jammed packings~\cite{Zachary2011,Dreyfus2015}. Since the discovery of the concept of hyperuniformity $15$ years ago~\cite{Torquato2003}, its realizations were found in numerous physical, biological, materials, and mathematical systems, and include disordered ground states, glass formation, jamming, Coulomb systems, spin systems, self-organization, retinal photoreceptor cells, and more (for a recent review, see Ref.~\cite{Torquato2018}).

Hyperuniformity thus provides a unified means to classify and structurally characterize crystals, quasicrystals, and special disordered point configurations~\cite{Torquato2018,Torquato2003}, and is tightly connected to a specific form of the structure factor of these systems. Despite this, the concept of hyperuniformity on the sphere remains by and large unexplored. The first attempts to define hyperuniformity on the sphere have been made only very recently by Brauchart et al.~\cite{Brauchart2018,Brauchart2018b}, who have introduced three different criteria for the behaviour of the number variance on the sphere that should be characteristic of hyperuniformity, and later by Meyra et al.~\cite{Meyra2018}, who first demonstrated different angular dependence scaling laws of the number variance. These attempts, however, do not relate the criteria for the behaviour of the number variance to the behaviour of the structure factor of particle distributions on the sphere, which itself has only recently started to be employed as a tool for analysis~\cite{Franzini2018}. Moreover, it remains unclear how the three different criteria of hyperuniformity on the sphere can be applied to finite systems encountered in practice, and characterization of finite particle systems and their relation to hyperuniformity requires further elucidation. Motivated in part by these developments, we set out to formulate a general notion of hyperuniformity for spherical systems.

In this work, we introduce the spherical structure factor of particle distributions on the sphere in the context of hyperuniformity and derive a general form of the number variance on the sphere. In particular, by relating the structure factor to the multipole expansion of particle distributions we show that the spherical structure factor of ordered spherical distributions has a gap at low multipole numbers which translates to a hyperuniform behaviour of the number variance. This provides a spherical analogue to the hyperuniformity criterion in Euclidean space that the structure factor vanishes in the limit of small wave vector magnitudes. Importantly, we show how the notion of hyperuniformity applies to systems with a finite number of particles, circumventing the need for a thermodynamic limit. We study several well-known particle distributions on the sphere, and we show that the gap in the spherical structure factor at low multipole numbers is independent of the underlying details of particle potential and seemingly stems simply from the presence of a minimum distance between the particles. Finally, we demonstrate for the first time that the number variance follows a universal form that can serve to derive order parameters that quantify the order of a particle distribution. These parameters are obtained easily by fitting and can be used to detect hyperuniformity in scale-free spherical distributions and quantify phase transitions into the hyperuniform state for distributions under variations of interaction length scales.

\section{Spherical structure factor}

We start with an arbitrary distribution of $N$ particles on a unit sphere, which can be characterized solely by the particle positions in spherical coordinates $\Omega_k=(\vartheta_k,\varphi_k)$, $k=1,\ldots,N$. The joint density (distribution) of the $N$ particles can be thus written as
\begin{equation}
\label{eq:rho}
\rho(\Omega)=\sum_{k=1}^N\delta(\Omega-\Omega_k),
\end{equation}
where $\delta$ is the Dirac delta function. This density can be further expanded in terms of multipole moments~\cite{ALB2018b},
\begin{equation}
\rho(\Omega)=\sum_{l=0}^\infty\sum_{m=-l}^l\rho_{lm}\,Y_{lm}(\Omega),
\end{equation}
where $\rho_{lm}$ are the multipole coefficients and $Y_{lm}(\Omega)$ are the spherical harmonics. The multipole expansion enables us to obtain the corresponding {\em multipole magnitudes}
\begin{equation}
\label{eq:Sl}
Q_l=\sqrt{\frac{4\pi}{2l+1}\sum_{m}|\rho_{lm}|^2},
\end{equation}
which have been previously shown to also have the form~\cite{ALB2018b}
\begin{equation}
\label{eq:xl}
Q_l=\left[N+2\sum_{k>t}P_l(\cos\gamma_{kt})\right]^{1/2}.
\end{equation}
The monopole moment $Q_0$ is simply proportional to the total number of the particles in the distribution $N$. Here, $P_l(x)$ are the Legendre polynomials and $\gamma_{kt}$ is the (great-circle) spherical distance between two particles located at $\Omega_k$ and $\Omega_t$, respectively. Following Brauchart et al.~\cite{Brauchart2018}, we now introduce the {\em spherical structure factor} of an $N$-particle distribution as
\begin{equation}
\label{eq:sl}
S_N(\ell)=\frac{1}{N}\sum_{k,t=1}^NP_\ell(\cos\gamma_{kt}),
\end{equation}
from where we immediately see the clear relation between the spherical structure factor and the multipole magnitudes,
\begin{equation}
S_N(\ell)=\frac{Q_\ell^2}{N}.
\end{equation}
We note in addition that similar measures have been used before in the studies of dynamics of particles interacting on the sphere in the form of a time-dependent scattering function~\cite{Vest2014,Vest2018}, whose static form we can now connect with the corresponding multipole moments.

An alternative definition of the spherical structure factor was recently presented by Franzini et al.~\cite{Franzini2018}. They use total (two-body) correlation function on the sphere $h_N(\Omega,\Omega')$~\cite{Franzini2018,Tarjus2011}:
\begin{equation}
\label{eq:pair}
\rho^2h_N(\Omega,\Omega')=\langle\rho(\Omega)\rho(\Omega')\rangle-\rho^2-\rho\delta(\Omega-\Omega'),
\end{equation}
and expand it in terms of multipole coefficients $h_N(\ell)$,
\begin{equation}
h_N(\ell)=\int h_N(\Omega,\mathbf{0})\,Y_{\ell 0}(\Omega)\,\mathrm{d}\Omega,
\end{equation}
where $\Omega'=\mathbf{0}$ is the position of the north pole. In this way, Franzini et al.~\cite{Franzini2018} derive the spherical structure factor as
\begin{equation}
\label{eq:sl2}
S_N(\ell)=1+\rho h_{N}(\ell),
\end{equation}
where $\rho\equiv\langle\rho(\Omega)\rangle=N/4\pi$. This definition together with Eq.~\eqref{eq:sl} connects the pair correlation function---which has previously been used to great extent to, for instance, analyze in detail the statistical geometry of hard particles interacting on the sphere~\cite{Giarritta1992,Giarritta1993,Fantoni2012}---with the spherical structure factor defined through the power spectrum of the multipole expansion of the particle density distribution. We note that for Eq.~\eqref{eq:pair} to be valid, large enough systems should be considered so that the number of particles can be considered as a continuous variable and explicit $1/N$ corrections can be neglected~\cite{Tarjus2011}; this restriction comes about due to the compactness of the sphere. What is more, $h_{N}(\ell)$ generally depends on $N$ while the density $\rho$ cannot be varied without changing $N$, so the relation in Eq.~\eqref{eq:sl2} does not reflect the complexity of the $N$-dependence.

Taken together, the definitions of the spherical structure factor in Eqs.~\eqref{eq:sl} and~\eqref{eq:sl2} can be considered as the spherical equivalent of the definition of the structure factor in Euclidean space~\cite{Torquato2018}, and we will show later on how they can be used to derive the criteria for hyperuniformity on the sphere. On the sphere, it is the multipole number $\ell$ that plays the role of the wave vector magnitude $|{\bf k}|$ from Euclidean space. Note that in Euclidean space, translational and rotational invariance are separate and some ordered structures, such as crystals and quasicrystals, may have preferred directions in space, which is mirrored in a directionally-dependent structure factor. Translational invariance on a sphere implies isotropy (in an average way, given the consideration that it is a finite system with mandatory lattice defects), and only the total contribution of each multipole plays a role.

\begin{figure*}
\begin{center}
\includegraphics[width=\textwidth]{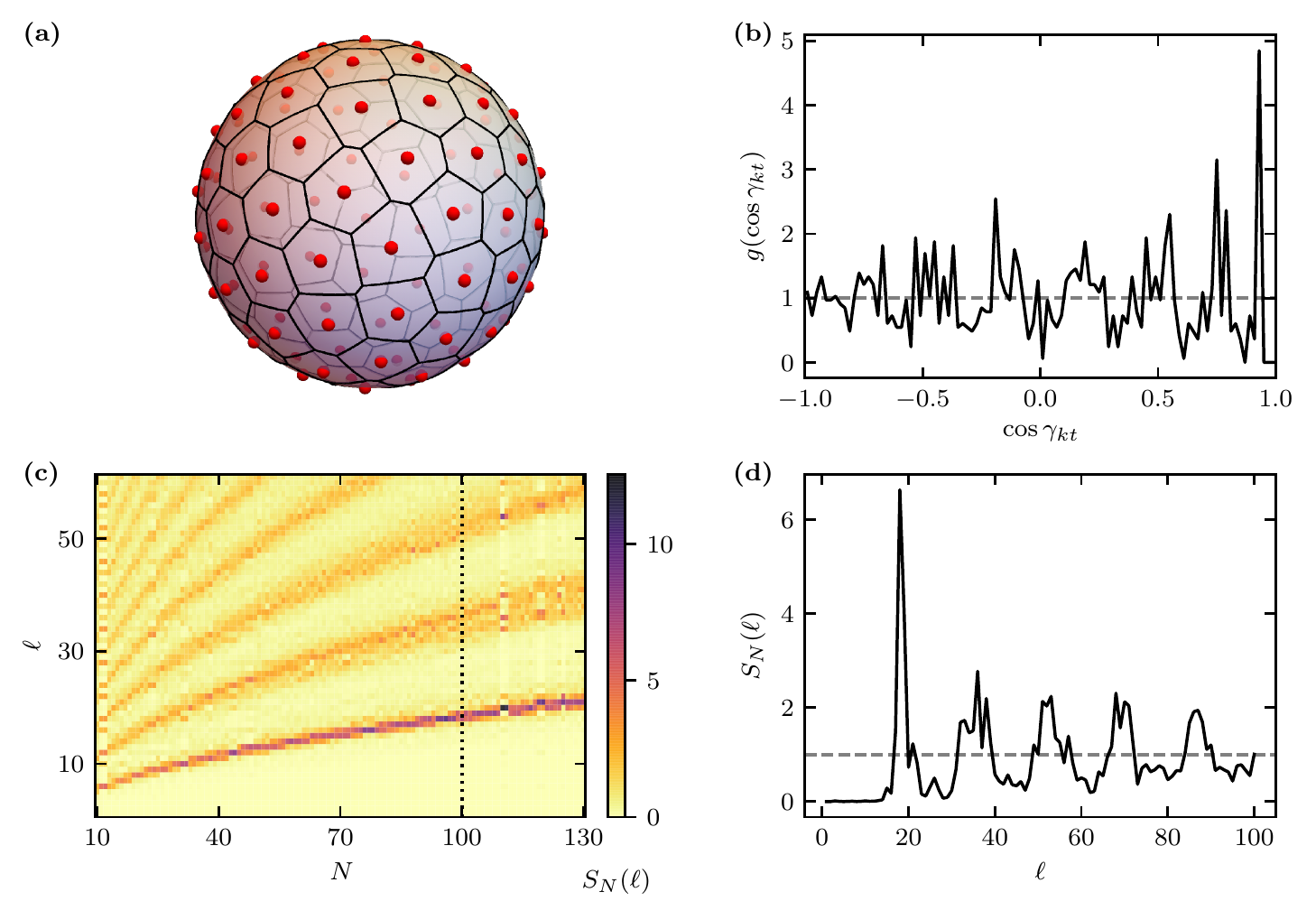}
\end{center}
\caption{{\bf (a)} Distribution of $N=100$ particles (red points) obtained as a solution of the Tammes problem---hard-disk packing on the sphere. Also shown is the Voronoi tesselation of the distribution (black lines). {\bf (b)} Pair distribution function $g(\cos\gamma_{kt})$ of the Tammes particle distribution from panel (a). {\bf (c)} Heatmap of the spherical structure factor $S_N(\ell)$ in the $(\ell,N)$ plane for distributions of particles based on the solutions of the Tammes problem. Dotted line shows the spherical structure factor of a distribution with $N=100$ particles [panel (a)], which is also shown separately in panel {\bf (d)}. Dashed gray lines in panels (b) and (d) show the pair distribution function and spherical structure factor of a uniform random distribution with $N=100$ particles.}
\label{fig:1}
\end{figure*}

To illustrate the spherical structure factor introduced in Eq.~\eqref{eq:sl}, we apply it to particle distributions resulting from solutions of the Tammes problem of hard-disk packing on the sphere~\cite{Saff1997} (obtained from N.\ Sloane's Spherical codes repository~\cite{SloaneWEBa}). Figure~\ref{fig:1}a shows the solution of the Tammes problem for $N=100$ particles, and Fig.~\ref{fig:1}b shows the corresponding pair correlation function $g(\cos\gamma)$ of the distribution (note that $g(\cos\gamma)=h(\cos\gamma)+1$ still holds). The pair correlation function naturally exhibits a gap at small distances ($\cos\gamma\approx1$) as a result of the hard-disk repulsion. In the spherical structure factor of the distribution, shown in Fig.~\ref{fig:1}d, a gap is present at low multipole numbers $\ell$. Figure~\ref{fig:1}c shows the structure factor in the $(\ell,N)$ plane, and it is clear that, with the exception of scaling with $\ell$, the structure factor has the same form for any number of particles in the distribution. Importantly, the low-$\ell$ gap in the structure factor persists for all $N$, and it can be shown for spherical hexagonal lattices to scale as $\ell_0\propto\sqrt{N}$, where $\ell_0$ is the position of the first peak in the structure factor (see Appendix~\ref{sec:appA}).

We also note here that, while depicted in Fig.~\ref{fig:1}d with a continuous line, the spherical structure factor is defined only for integer (non-negative) values of multipole numbers $\ell$. This is in contrast to the structure factor in Euclidean space, where the wave vector is (in unbounded space) continuous. As a consequence of this, a strict limit $\ell\to0$ of the spherical structure factor does not exist; however, we can clearly still determine whether the spherical structure factor {\em vanishes} at low $\ell$. We will only consider well-behaved distributions without extraneous resonant behaviour at very low multipoles, so that at $\ell\ll\sqrt{N}$, the structure factor can be approximated with a \emph{continuous function} for which the limit $\lim_{\ell\to 0}S_N(\ell)\equiv \overline{S}_N(0)$ exists. At higher multipole numbers, which describe features that are small with respect to the radius of curvature, the Euclidean limit is naturally approached.

As we are dealing with point distributions, the spherical structure factor possesses additional properties that will be used in our further calculations. The Dirac delta function is not band-limited and the high-frequency limit of the structure factor is
\begin{equation}
\label{eq:sn_limit}
\lim_{\ell\to\infty}S_N(\ell)=1,
\end{equation}
where we have already adopted a continuous approximation for $S_N(\ell)$. Additionally, the normalization of the structure factor is set by the number of particles through the Parseval's theorem. Though the Dirac delta function distribution is not square-integrable, normalization can be expressed relative to the high frequency limit,
\begin{equation}
\label{eq:parseval}
\int_0^\infty\left[S_N(\ell)-1\right]\ell\,{\rm d}\ell=0,
\end{equation}
which will prove important to ensure consistently normalized analytical models.

\section{Cap number variance}

In Euclidean space, number variance is a measure of spatial density fluctuations in a system and is defined as the variance of the number of particles contained in a (hyper)spherical window with radius $R$, averaged over all possible positions of the window~\cite{Torquato2003,Torquato2018}. Extending the notion of number variance to the surface of a sphere brings with it certain restrictions not present in Euclidean space. Systems of finite $N$ are more important here, and they have to be considered independent of the size of the sphere, since density is no longer a good parameter~\cite{Hill1994,Post1986}. Moreover, there is no limit of large observation windows, $R\to\infty$, in contrast to Euclidean space, nor does averaging over window positions collect an infinite number of independent samples as the system size is finite. We show next that it is nonetheless possible to define {\em cap number variance}, defined on the sphere as the variance of the number of particles in a series of spherical caps, and derive the limiting forms of its behaviour based on its connection to the spherical structure factor. This analysis will then allow us to introduce and study the notion of hyperuniformity on the sphere and classify its behaviour.

We define a spherical cap $C(\Omega_C,\theta)$, centred on $\Omega_C$ and with an opening angle $\theta$, as the set of points on the sphere whose spherical distance to $\Omega_C$ is at most $\theta$. This is the most natural choice of the observation window on the sphere, providing an analogue to the (hyper)spherical window in the Euclidean case~\cite{Torquato2003}. Of our interest is the number of particles contained in a spherical cap,
\begin{equation}
N(\theta)=\int_{\mathbb{S}^2}\rho(\Omega)\,C(\Omega_C,\theta)\,\mathrm{d}\Omega,
\end{equation}
and the corresponding cap number variance,
\begin{equation}
\label{eq:var1}
\sigma_N^2(\theta)= \left\langle\vert N(\theta)\vert^2\right\rangle-\vert\left\langle N(\theta)\right\rangle\vert^2,
\end{equation}
where the averages are taken over all possible positions (centres) of the spherical cap, $\Omega_C\in{\mathbb{S}}^2$. The finite size of the system induces a symmetry in the cap variance---the numbers of particles outside and inside each spherical cap are constrained to add up to $N$, so their cap variances are the same: $\sigma^2_N(\theta)=\sigma^2_N(\pi-\theta)$.

While Eq.~\eqref{eq:var1} allows us to numerically determine the cap number variance of any particle distribution, we can also express it solely in terms of the spherical structure factor  of the distribution. Derivation, detailed in Appendix~\ref{sec:appB}, yields
\begin{equation}
\label{eq:var}
\sigma_N^2(\theta)=\frac{N}{4}\sum_{\ell=1}^\infty S_N(\ell)\,\frac{\left[P_{\ell+1}(\cos\theta)-P_{\ell-1}(\cos\theta)\right]^2}{2\ell+1}.
\end{equation}
Observe that this form of the cap number variance combines the spherical structure factor of a distribution with the multipole expansion (spherical analogue of the Fourier transform) of the ``volume'' function on the sphere, in analogy to the results derived by Torquato and Stillinger~\cite{Torquato2003} in Euclidean space. The factor involving the Legendre polynomials stands in for the Fourier transform of the scaled intersection volume in Euclidean case (Eq.~(73) in Ref.~\cite{Torquato2018}). At the same time, Eq.~\eqref{eq:var} presents a derivation and formulation of cap number variance alternative to the one given by Eq.~(10) in Ref.~\cite{Brauchart2018}.

\subsection{General form of cap number variance}
\label{ssec:gen_form}

The spherical structure factor of a uniform random distribution satisfies $S_N(\ell)=1$ $\forall\ell\geqslant1$ (alternatively, this means that $Q_\ell=\sqrt{N}$ $\forall\ell\geqslant1$)~\cite{ALB2018b,Brauchart2018}. In this case, Eq.~\eqref{eq:var} can be shown to converge pointwise to
\begin{equation}
\label{eq:var_rand}
\sigma_N^2(\theta)=\frac{N}{4}\sin^2\theta.
\end{equation}
This is the same dependence as obtained by considering completely independent particle placements on the sphere, resulting in a binomial number variance, $\sigma_N^2(\theta)=N p\,(1-p)=N\Sigma_C(1-\Sigma_C)=N\sin^2\theta/4$, where $\Sigma_C=(1-\cos\theta)/2$ is the normalized area of a spherical cap. The derived expression for cap number variance, Eq.~\eqref{eq:var}, thus correctly reduces to the form of a uniform random distribution when the appropriate form of the spherical structure factor is considered. A notable difference from the Euclidean case is that the cap number variance of a random distribution does not scale as the area of the spherical cap but scales instead as the square of its circumference, which starts decreasing for $\theta>\pi/2$.

To arrive at a general expression of the cap number variance for an arbitrary form of the spherical structure factor, we construct a Green's function---a response to a single normalized peak in the structure factor at $\ell=\ell_0$. In the large-$\ell$ approximation, the impulse contribution to the cap number variance yields (see Appendix~\ref{sec:appC} for a detailed derivation):
\begin{equation}
\label{eq:var_sin}
\sigma_{N,\ell=\ell_0}^2(\theta)\asymp\frac{N}{4}\frac{2}{\pi\,\ell_0^2}\,\sin\theta+\xi(\ell_0,\theta)
\end{equation}
where $\xi(\ell_0,\theta)$ is an oscillatory remainder term that reflects the periodic behaviour of the pair distribution function. Such an oscillatory remainder is expected for ordered (crystalline) distributions with sharp peaks in their structure factor---such as the Tammes distribution, shown in Fig.~\ref{fig:1}. In most cases, the oscillatory contribution is either dominated by the first peak of the structure factor (see Fig.~\ref{fig:C1} in Appendix~\ref{sec:appC}), or is not present at all if the structure factor has no sharp resonances. The oscillatory remainder term $\xi(\ell_0,\theta)$ thus represents a higher-order contribution and will be disregarded in our further analysis.

The remarkable result of Eq.~\eqref{eq:var_sin} is a universal angular dependence of the cap number variance $\sigma_N^2\propto \sin \theta$, which is valid as long as the structure factor has a gap around $\ell\to0$, required by the approximations we undertook in the derivation (Appendix~\ref{sec:appC}). As $\sin\theta$ is the normalized circumference of the spherical cap, we can already draw a parallel with the Euclidean case, where the number variance of hyperuniform distributions scales with the circumference of the spherical observation window.

For the special case of the Tammes, Thomson, and other ordered distributions, the structure factor can be modeled as a series of equidistant peaks (see Fig.~\ref{fig:1}c). The peaks are spaced by $\ell_0\approx\pi\sqrt{N}/\sqrt{3}$, derived under assumption of a spherical hexagonal lattice (Appendix~\ref{sec:appA}), and their magnitudes are equal to $\ell_0$, set by the normalization condition in Eq.~\eqref{eq:parseval}. Summation of the Green's function [Eq.~\eqref{eq:var_sin}] then yields an approximation for the cap number variance
\begin{equation}
\label{eq:var_gapped}
\sigma_N^2(\theta)\approx\frac{\sqrt{N}}{4\sqrt{3}}\sin\theta.
\end{equation}
More details of the derivation are given in Appendix~\ref{sec:appC}.

If the spherical structure factor is well-behaved enough at low multipole numbers to determine a limiting value $\overline{S}_N(0)$ (and assuming a continuous approximation of the structure factor), we can split it into a constant part, with its contribution to the number variance given by Eq.~\eqref{eq:var_rand}, and a variable part with a vanishing value around small $\ell$, whose contribution to number variance can be expressed through the integral of the Green's function [Eq.~\eqref{eq:var_sin}]. The cap number variance then asymptotically expands into two terms that scale as circumference squared and circumference of the cap, respectively:
\begin{eqnarray}
\label{eq:expansion}
\nonumber\sigma^2_N(\theta)&\asymp&\frac{N}{4}\Bigg[\overline{S}_N(0)\sin^2\theta\\
&&\phantom{\frac{N}{4}}+\frac{2}{\pi}\sin\theta\int_0^\infty\frac{S_N(\ell)-\overline{S}_N(0)}{\ell^2}\,{\rm d}\ell\Bigg].
\end{eqnarray}
This expression is suitable for smooth structure factors that can be modeled by analytical functions, and shows that the prefactor of the $\sin\theta$ term will vary between different distributions.

Regardless of the details of the spherical structure factor, the cap number variance of $N$ particles can be expressed as a linear combination of two terms that scale as Eq.~\eqref{eq:var_rand} and Eq.~\eqref{eq:var_gapped}, respectively, with unknown numerical factors that can be fitted to empirically obtained cap number variance:
\begin{equation}
\label{eq:var_fit}
\sigma_N^2(\theta)\asymp \frac{A_N}{4}N\sin^2\theta+\frac{B_N}{4 \sqrt{3}}\sqrt{N}\sin\theta + \xi(\theta).
\end{equation}
The first term, proportional to the square of the cap's circumference, stems from the contribution of the randomness (disorder) in the distribution and scales with the number of particles $N$. The second term is proportional to the circumference of the spherical cap---observation window---and scales with $\sqrt{N}$. Judging by the analytical expansion of Eq.~\eqref{eq:expansion}, the coefficient $B_N$ will generally decrease as $A_N$ increases, but will also depend on the detailed shape of the structure factor. Oscillatory higher-order corrections $\xi(\theta)$, which are expected to be present in crystalline-like distributions just like in the Euclidean case, present a minor contribution that, due to its oscillatory nature, does not tend to decrease the quality of the fit (for details, see Appendix~\ref{sec:appC}).

\subsection{Examples: spherical structure factor and cap number variance}
\label{sec:expl}

\begin{figure}[!b]
\begin{center}
\includegraphics[width=\columnwidth]{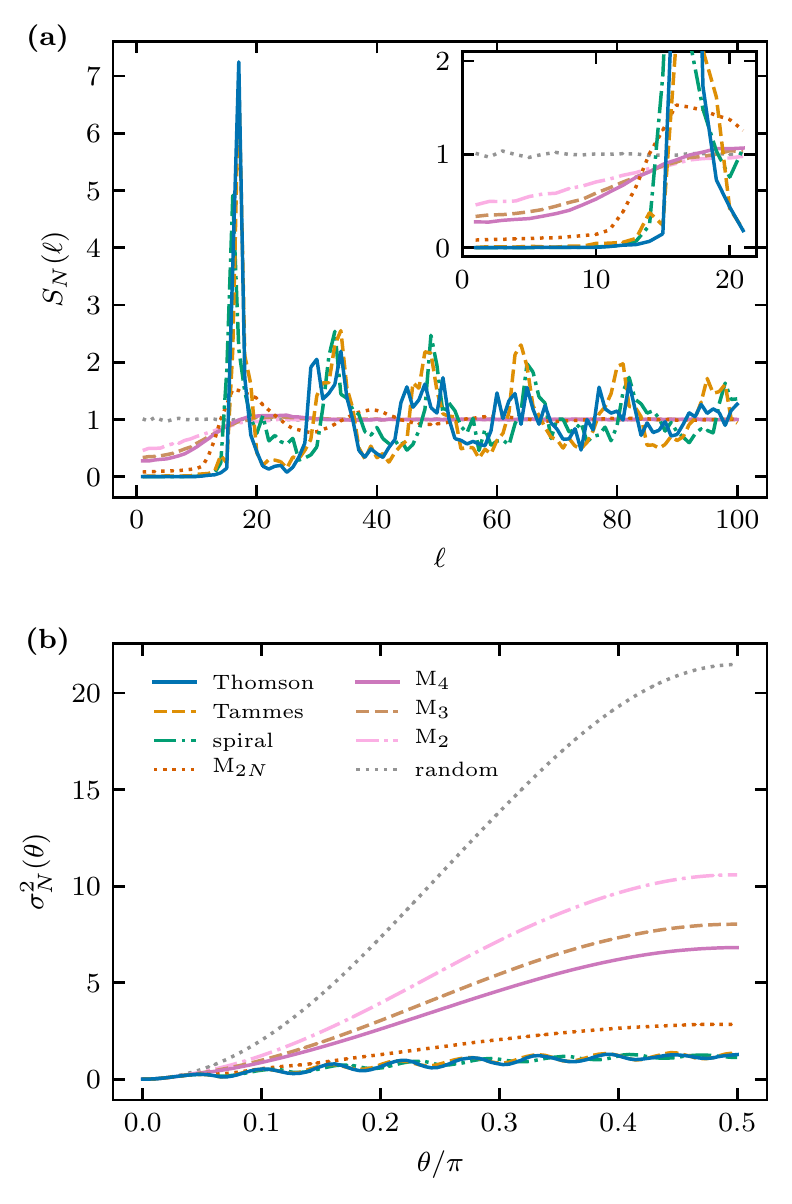}
\end{center}
\caption{{\bf (a)} Spherical structure factor $S_N(\ell)$ and {\bf (b)} cap number variance $\sigma_N^2(\theta)$ for eight different types of distributions of $N=85$ particles. The inset in panel (a) shows a zoomed-in region at low multipole numbers $\ell$. The values of $\sigma_N^2(\theta)$ are averaged over $2\times10^4$ cap sequences at each $N$ and $\theta$, and those of randomly-generated distributions are additionally averaged over $1000$ different configurations [in both panels (a) and (b)]. The spherical distributions depicted in panel (a) correspond to those in panel (b) with the plot styles specified in the legend of  panel (b).}
\label{fig:2}
\end{figure}

Next, we demonstrate the different forms the spherical structure factor can take on for different types of distributions, and we show how these forms translate into differences in their respective cap number variances. We do this for several different types of distributions, described in detail in Appendix~\ref{sec:appD}. These include both ordered distributions (solutions of the Thomson and Tammes problems, and generalized spiral nodes), as well as randomly-generated distributions (uniform random distribution and distributions generated utilizing Mitchell's algorithm, which allows each particle $K$ trials to be randomly repositioned upon addition, retaining the position furthest away from the particles already placed; the latter distributions will be denoted as M$_K$). Since we are interested in finite systems, we mostly limit our analysis to small system sizes, $N\leqslant150$. The extension to larger $N$ is trivial for stochastic systems due to the nature of their generation. For ordered systems---namely, for Thomson and spiral distributions---we also include distributions with up to $N\sim500$ particles in order to verify that the results of our work remain valid for large system sizes. Appendix~\ref{sec:appD} lists the exact range of $N$ used for each analyzed distribution.

Figure~\ref{fig:2}a shows the spherical structure factor for eight different types of these distributions containing $N=85$ particles, and we can immediately see that different distributions can be distinguished by the form of their structure factor. Ordered distributions (Thomson, Tammes, and spiral) have a crystalline-like structure factor with a pronounced first peak followed by an equidistant series of broader ones. This seems to be a generalization of the structure of the multipole moments of spherical distributions with high symmetry, where only select multipole numbers $\ell$ are permitted~\cite{Lorman2007,ALB2013a}. Mitchell distributions M$_K$ with a small number of allowed trials $K$ have, on the other hand, a structure factor which quickly approaches the one characteristic of the uniform random distribution, $S_N(\ell)=1$ $\forall\ell\geqslant1$. The number of allowed trials also appears to determine the form of their structure factor at low $\ell$ (inset of Fig.~\ref{fig:2}a), which---on average---gets smaller, the more trials a particle has to reposition itself. When we allow the number of trials to scale linearly with $N$, as is the case for M$_{2N}$ distribution, we observe a jammed-like distribution with a broad and flat-bottomed gap and a very small value of $S_N(\ell)$ at low $\ell$. Accordingly, the form of the spherical structure factor at low $\ell$ appears to indicate the deviation of a distribution from a completely random one, and the structure factor at low $\ell$ can be observed to completely vanish for ordered distributions.

In Fig.~\ref{fig:2}b we show the cap number variance of the same distributions, calculated according to Eq.~\eqref{eq:var1} by counting the number of particles in $2\times10^4$ randomly placed spherical caps for each cap opening angle $\theta$. The differences between the distributions are now presented in a complementary way to Fig.~\ref{fig:2}a, as the more ordered distributions have an overall much smaller cap number variance, a less pronounced curve, and are modulated by oscillations indicating an underlying ordered lattice structure. Based on our considerations in Sec.~\ref{ssec:gen_form}, we can predict that the cap number variance of different distributions scales differently with both $\theta$ and $N$, which we quantify in the next Section to extract hyperuniformity information of these spherical distributions.

\section{Hyperuniformity on the sphere}

Structure factor, defined through a Fourier transform of the pair distribution function, is an essential component when determining hyperuniformity of a distribution in Euclidean space: hyperuniform states of matter possess a vanishing structure factor at small wave vector magnitudes, $\lim_{|{\bf k}|\to0}S({\bf k})=0$~\cite{Torquato2003,Torquato2018}. At the same time, hyperuniformity of a system means that its spatial density fluctuations are suppressed at long wavelengths---large length scales. By counting the number of particles in observation windows of size $R$ in $d$-dimensional space, the variance of this number scales as $R^d$ for completely random particle distributions. In hyperuniform distributions, on the other hand, the number variance scales with a different power of $R$. Thus, hyperuniform systems in Euclidean space can be divided into three classes, based on their behaviour at $R\to\infty$~\cite{Torquato2018}: {\em (i)} $R^{d-1}$ (e.g., crystals and some quasicrystals); {\em (ii)} $R^{d-1}\ln R$ (e.g., some quasicrystals, fermionic point processes, or maximally jammed random packings); and {\em (iii)} $R^{d-\alpha}$, $0<\alpha<1$ (e.g., some perfect glasses or perturbed lattices).

To translate the hyperuniformity concepts to systems on the sphere, parallels can be drawn between Euclidean and spherical structure factors, and between the number variance and the spherical cap number variance. However, there are also fundamental differences that we have already touched on briefly: on the sphere, finite system size prevents a proper limit of a large number of particles, and the multipole number (wave vector) in the spherical structure factor is discrete instead of continuous. A consistent definition of hyperuniformity must thus reconcile these differences, and we approach the challenge of finite systems in the next Subsection.

We have already shown that ordered distributions on the sphere possess a spherical structure factor which has a gap at low multipole numbers, meaning that it is possible to talk about a vanishing structure factor in the limit $\lim_{\ell\to0}S_N(\ell)=0$ while still keeping in mind its discrete nature. The different forms of the structure factor clearly translate into different forms of the corresponding cap number variance---this is observed both in different distributions we have considered, which span from completely ordered to completely random (Fig.~\ref{fig:2}), as well as in the general form of the number variance derived by considering a response to an impulse function or a gap in the structure factor, which predicts the two dominant types of behaviour [Eq.~\eqref{eq:var_fit}]. The description is thus sufficiently general to formulate a definition of hyperuniformity without referring to specific details of model distributions.

\subsection{Scaling behaviour in finite systems}

Thus far, we have kept the subscript $N$---denoting the number of particles in the system---for both the spherical structure factor $S_N(\ell)$ as well as the cap number variance $\sigma_N(\theta)$ and the coefficients $A_N$ and $B_N$ of its general form. Unlike the Euclidean case where it is possible to talk about these quantities in the limit of $N\to\infty$, most configurations of interest on the sphere involve a finite number of particles. What is more, the particle density $\rho\propto N/r^2$ can be varied by changing either the number of particles $N$ or the size of the sphere $r$, which have to be treated as two independent parameters~\cite{Hill1994,Post1986}. Scaling the sphere radius is equivalent to scaling the interparticle interaction, which results in different behaviour depending on the number of characteristic length scales:
\begin{enumerate}
\item {\bf Scale-free systems:} $N$ is the only parameter of the system. Examples include Tammes distribution, all power-law interactions including the logarithmic and Coulomb (Thomson) distributions, uniform random distribution, Markovian stochastic distributions, generalized spiral distributions, and honeycomb lattice distributions.
\item {\bf Systems with one or more length scales}: both $N$ and $\rho$ can be varied independently. Examples include particles interacting through Lennard-Jones potential, GEM-4 potential, hard-core--soft-shoulder interactions, \ldots
\end{enumerate}

In scale-free systems there is no additional density parameter to be varied at fixed $N$---there is no ``compression'' response, as rescaling the size does not change the distribution. No phase transition is possible. In systems with a length scale, on the other hand, $N$ can be varied together with the size of the system, keeping the density constant. Dependence on $N$ reveals the effect of curvature and finite size, and depends on the type of interaction. In the limit of $N\to\infty$, the curvature effects vanish and we recover the Euclidean case, and $A_N$ and $B_N$ will converge to a constant. If the density is increased at constant $N$, the system will exhibit interaction-specific phase behaviour and can consequently exhibit a phase transition between hyperuniform and non-hyperuniform states. Study of phase behaviour of such systems is beyond the scope of this article. We will thus further focus only on particle distributions which do not possess an intrinsic length scale---the distributions introduced in Sec.~\ref{sec:expl} and detailed in Appendix~\ref{sec:appD}---and comment on systems with intrinsic length scales in the Discussion.

\subsection{Classification of hyperuniform distributions on the sphere}

\begin{figure}[!tb]
\begin{center}
\includegraphics[width=\columnwidth]{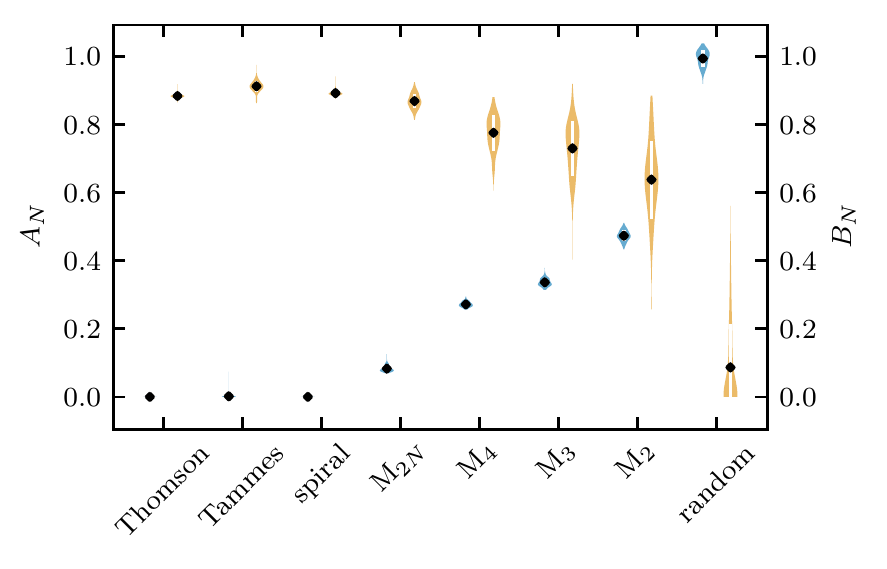}
\end{center}
\caption{Violin plot of the coefficients $A_N$ and $B_N$ of the generalized form of cap number variance, Eq.~\eqref{eq:var_fit}, for different types of distributions, fitted over the entire range $\theta\in[0,\pi/2]$. The coefficient pertaining to the scaling of the number variance as a random distribution, $A_N$, is shown in blue on the left part of each distribution, and the coefficient pertaining to the scaling of the number variance as the circumference of the cap, $B_N$, is shown in yellow on the right part of each distribution. The width of each entry in the violin plot shows a distribution of the fitted coefficients for a range of $N$ (see Appendix~\ref{sec:appD}), with the central black symbols denoting the mean and the white bars denoting the corresponding standard deviation.}
\label{fig:3}
\end{figure}

We can now write down the {\em criteria for hyperuniformity} of particle distributions on the sphere, analogous to those in Euclidean space: a hyperuniform distribution possesses a spherical structure factor which vanishes at low multipole numbers $\ell$ (inset of Fig.~\ref{fig:2}a), similar to a limit of the form $\lim_{\ell\to0}S_N(\ell)=0$. This, in turn, translates into cap number variance which scales as the circumference of the cap, $\sigma_N^2(\theta)\propto\sin\theta$, and has a vanishing factor $A_N=0$ in its generalized form of Eq.~\eqref{eq:var_fit} (Fig.~\ref{fig:3}). For practical purposes, the structure factor needs not vanish completely and a reasonably small $A_N$ still allows hyperuniform behaviour. Contrary to the Euclidean case, the non-hyperuniformity limit is not given by the scaling of the cap number variance as the surface of the cap $\Sigma_C$, but is instead proportional to $\Sigma_C(1-\Sigma_C)$, scaling as the square of the circumference of the cap, $\sigma_N^2(\theta)\propto\sin^2\theta$.

For hyperuniform distributions, the factor $B_N$ holds additional information about the distribution---as shown in Appendix~\ref{sec:appC}, a regular crystalline-like distribution has a higher $B_N$ value compared to a disordered jammed-like distribution with a finitely wide gap but no distinct peaks in the structure factor. Spherical cap number variance of any distribution can be calculated either directly from the positions of the particles or, alternatively, from the spherical structure factor. This formulation does not require a thermodynamic limit and can be applied to individual measured or generated distributions, as well as to canonical or grand-canonical ensembles of distributions.

To determine the scaling of the cap number variance more exactly, we can fit the generalized form of the number variance in Eq.~\eqref{eq:var_fit} to different distributions, thus obtaining their coefficients $A_N$ and $B_N$ which pertain to the two major forms of scaling of the cap number variance. The fits are taken over the entire range of $\theta\in[0,\pi/2]$ for each individual $N$ (see Appendix~\ref{sec:appD} for the ranges of $N$ considered), and the fitted coefficients are shown in Fig.~\ref{fig:3} in the form of a violin plot. Several things can be observed immediately: all of the ordered distributions (Thomson, Tammes, and spiral) have the factor $A_N$ equal to zero. This means that their cap number variance scales as the circumference of the cap---the observation window on the sphere---thus clearly indicating their hyperuniformity. A completely random distribution is on the opposite side of the spectrum, having $A_N=1$ and a vanishing coefficient $B_N$. (We note here that the spread in the values of $B_N$ is due to this term being subdominant when $A_N$ is large, making their fitted values less reliable.) Mitchell distributions, which allow a number of trial particle placements throughout their generating procedure, fall somewhere in between the two extremes, as they have both coefficients $A_N$ and $B_N$ non-zero. The more relocations a particle is permitted---and, consequently, the less of a role randomness plays in the generating procedure---the lower the $A_N$ and the higher the $B_N$ coefficient. Moreover, the fit hints at $A_N\sim 1/K$ for low number of trials in Mitchell M$_K$ distributions. When the number of allowed trials scales with $N$, as it does in the M$_{2N}$ distribution, the $A_N$ coefficient becomes very small, $A_N\lesssim0.1$, while the coefficient $B_N$ is comparable to those of the ordered distributions. Thus, the Mitchell family of distributions demonstrates a gradual transition between random and ordered structures, which is reflected directly in the parameters $A_N$ and $B_N$.

\begin{figure}[!b]
\begin{center}
\includegraphics[width=\columnwidth]{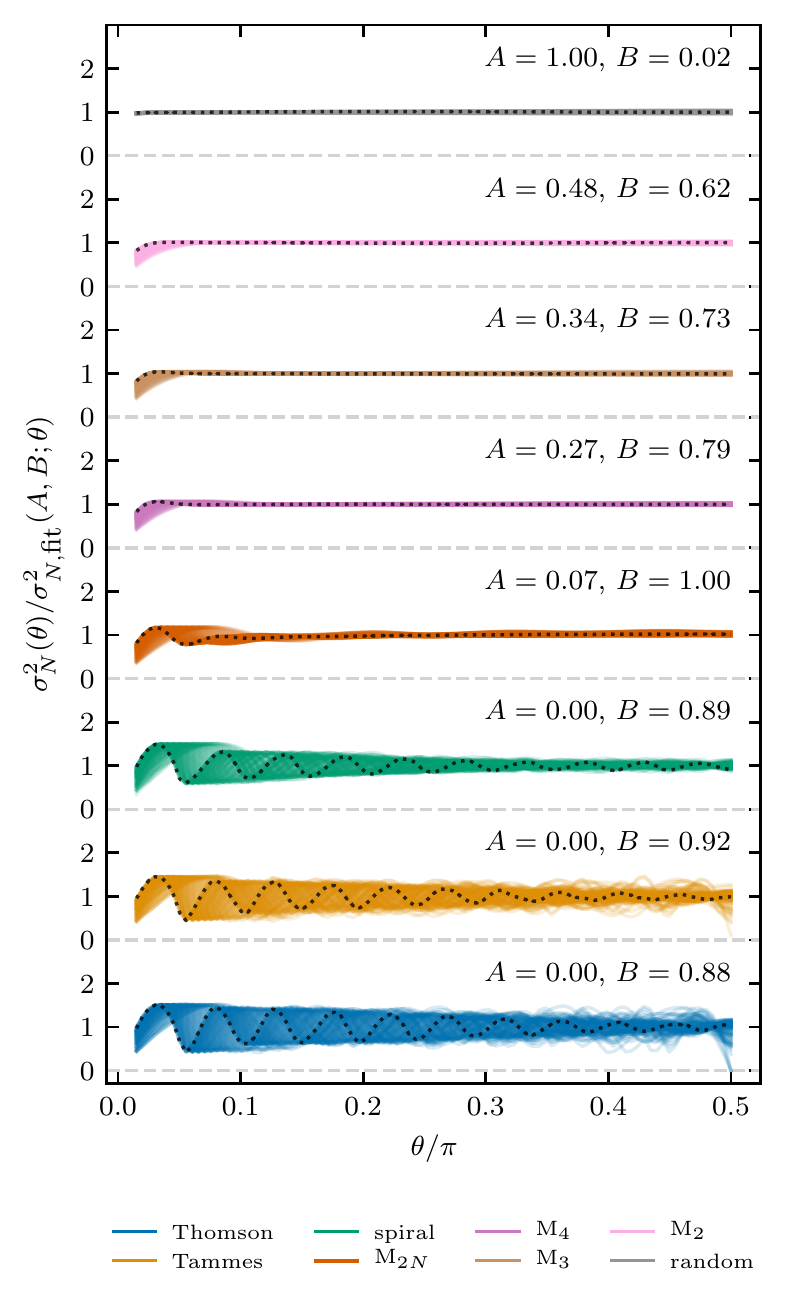}
\end{center}
\caption{Cap number variance $\sigma_N^2(\theta)$ of eight different types of distributions, scaled with the generalized form of number variance [Eq.~\eqref{eq:var_fit}] with fitted coefficients $A$ and $B$, denoted in the plot. The fits were performed simultaneously in the $(N,\theta)$ space over the entire range of $\theta\in[0,\pi/2]$ and the available range of $N$ for each distribution (see Appendix~\ref{sec:appD}). Overlaid are curves for $N\in[20,130]$, and the dotted black lines show the rescaled curves for $N=130$. Dashed gray lines show the value of $0$. The values of cap number variance are averaged over $2\times10^4$ cap sequences at each $N$ and $\theta$, and those of randomly-generated distributions are additionally averaged over $1000$ different configurations.}
\label{fig:4}
\end{figure}

The relatively small spread of the values of the coefficients $A_N$ and $B_N$ with respect to $N$ for scale-free distributions, as seen in Fig.~\ref{fig:3}, indicates that they are largely independent of $N$, as predicted for the simplest distributions by Eqs.~\eqref{eq:var_rand} and~\eqref{eq:var_gapped}. This leads us to fit the generalized form of the cap number variance simultaneously over the entire range of both $N$ and $\theta$, thus obtaining a single pair of coefficients $A$ and $B$ for each type of distribution, which should lead to a collapse of all their cap number variance curves. That this is indeed so is demonstrated in Fig.~\ref{fig:4}, which also shows the values of the fitted coefficients (note that they are virtually the same as the means of individually fitted coefficients in Fig.~\ref{fig:3}, with the exception of the coefficient $B$ of M$_{2N}$, which is now slightly larger, $B=1$ compared to $\langle B_N\rangle_N=0.87$). Distributions with crystalline-like order exhibit residual oscillations with near-vanishing amplitude at $\theta=\pi/2$ (lower three panels in Fig.~\ref{fig:4}). These oscillations can be attributed mainly to the first peak at $\ell=\ell_0$ with unit half-width and are proportional to $\sin((2\ell_0+1)\theta)$. For more details on the term describing the residual oscillations, see Appendix~\ref{sec:appC}.

Scale-free distributions can thus be characterized only by two parameters $A$ and $B$ which are independent of $N$ (Fig.~\ref{fig:4}). Hyperuniformity is determined by the parameter $A$, which can then be used to quantify order or disorder in a distribution. This is a global parameter and can also serve as a complement to the bond order parameter, especially when it comes to quantification of disorder~\cite{Mickel2013}.

\section{Discussion}

There are some necessary conditions for distributions to fall into the framework of hyperuniformity derived in this work. One main prerequisite for hyperuniformity is uniformity of the particle distribution in the sense of rotational isotropy, which is hard to rigorously define for a finite number of particles, especially for stochastic distributions. When large-scale deviations in particle density are present along the surface of the sphere, lower multipole components will exhibit behaviour which nullifies our condition that $S_N(\ell)-\overline{S}_N(0)$ should vanish for multipole numbers that are small compared to the scale of average interparticle distance (modelled by $\ell_0\propto \sqrt{N}$).

The choice of the method to determine hyperuniformity of a distribution depends on the size and format of the data. Cap number variance can be measured directly by randomly sampling a sufficiently large number of cap centres on the sphere, sorting the points according to the distance from the chosen centre, and computing the averages in Eq.~\eqref{eq:var1}. In this form, fitting Eq.~\eqref{eq:var_fit} to the form of the cap number variance reliably determines the hyperuniformity parameters $A$ and $B$. Alternatively, spherical structure factor can be computed from Eq.~\eqref{eq:sl}, for example, in which case we can directly observe the shape of the gap and obtain an estimate of $A$ from the lowest multiple terms $\ell$. In most cases, a limited lower range of the spectrum will suffice to estimate the hyperuniformity parameters. If we consider smooth ``particle'' distributions---i.e., not composed of point particles, but smeared out instead---the structure factor no longer obeys the limit $S_N(\ell)\asymp 1$ when $\ell\to\infty$, as smoothing reduces the importance of longer length scales~\cite{ALB2018b}. Discrete summation formulae cannot be used for continuous distributions where particle positions or even their number are undefined, but the structure factor can nonetheless be computed for the relevant low-$\ell$ range.

We have studied the concepts of hyperuniformity on the sphere only on distributions which possess no internal scale. We have shown that for these, the parameters $A_N$ and $B_N$ are independent of $N$. Our asymptotic form for the cap number variance [Eq.~\eqref{eq:var_fit}], put to the test in Fig.~\ref{fig:4}, provides a robust set of parameters $A$ and $B$. For a single distribution, these indicate the presence of hyperuniformity through the parameter $A$ and give more information of the degree and type of hyperuniform order through the parameter $B$. For deterministic distributions and those obtained from single measurements, cap number variance and the fitted parameters $A$ and $B$ are reliable quantifiers of hyperuniformity and degree of order. For stochastic distributions, individual realizations from the same ensemble follow a certain distribution in $A$ and $B$ parameters, as indicated by the violin plot in Fig.~\ref{fig:3}, which can be exploited to probe the properties of the ensemble or even differentiate between ensembles.

In systems with an intrinsic length scale, a phase transition into a hyperuniform state may be triggered by compression to critical density (similar to the Euclidean case), in which case the $N$ dependence will also be present in the variation of the $A_N$ and $B_N$ parameters. In such distributions, the vanishing of the $A_N$ parameter will coincide with the reduction of the structure factor to zero at low multipole numbers, creating a finitely wide gap. Examples of distributions with an intrinsic length scale include distributions of particles interacting through Lennard-Jones and Morse potentials~\cite{Vest2014,Paquay2016,Vest2018}, generalized exponential models (e.g., GEM-4)~\cite{Franzini2018}, or hard-core--soft-shoulder (HC-SS) potential~\cite{Julija2011}, and these interactions lead to the resulting distributions exhibiting numerous phases~\cite{Miller2011,Julija2011}. Certain interparticle potentials even lead to a formation of local clusters of particles which then form global crystalline-like structures~\cite{Franzini2018}. While a comprehensive discussion of these distributions requires a separate treatment, we have tested the hyperuniformity concepts on GEM-4 and HC-SS distributions, published previously in Refs.~\cite{Franzini2018} and~\cite{Julija2011}. The framework presented in this work remains applicable also there, with the difference that the fit coefficients are no longer independent of $N$, but can instead trace the phase transitions that occur when $N$ is varied, as the hard-coded $N$ and $\sqrt{N}$ scaling factors in Eq.~\eqref{eq:var_fit} only hold within a single ``phase''. To give them justice, these systems call for a detailed, separate investigation, as several parameters need to be explored simultaneously.

As mentioned in the Introduction, hyperuniformity on the sphere has also been considered very recently by Brauchart et al.~\cite{Brauchart2018,Brauchart2018b} and Meyra et al.~\cite{Meyra2018}. Brauchart et al.~\cite{Brauchart2018} have based their definition of hyperuniformity solely on the form of the cap number variance, wherefrom they have derived, in the limit of $N\to\infty$, three criteria for hyperuniformity based on the range and scaling of cap opening angles $\theta$ in question. Our definition thus presents several advantages: it is applicable also for finite, small $N$, and can be used simultaneously over the entire range of $\theta$. It furthermore connects a vanishing spherical structure factor with a cap number variance which depends only on the circumference of the spherical cap, in perfect analogy to the definition of hyperuniformity in Euclidean space. The limiting extremes of $\sin\theta$ and $\sin^2\theta$ scaling behaviour of the cap number variance were first observed very recently by Meyra et al.~\cite{Meyra2018}; we here show their linear combination describes the whole space of possibilities (notwithstanding residual oscillations that are particular to each distribution). Their superposition, however, can easily mimic a form of $\sigma_N^2(\theta)\propto\sin^\alpha\theta$, observed in Ref.~\cite{Meyra2018}. Such a form can be considered an empirical model but does not have a theoretical foundation, excepting specifically-constructed pathological cases.

\begin{figure}[!tb]
\begin{center}
\includegraphics[width=\columnwidth]{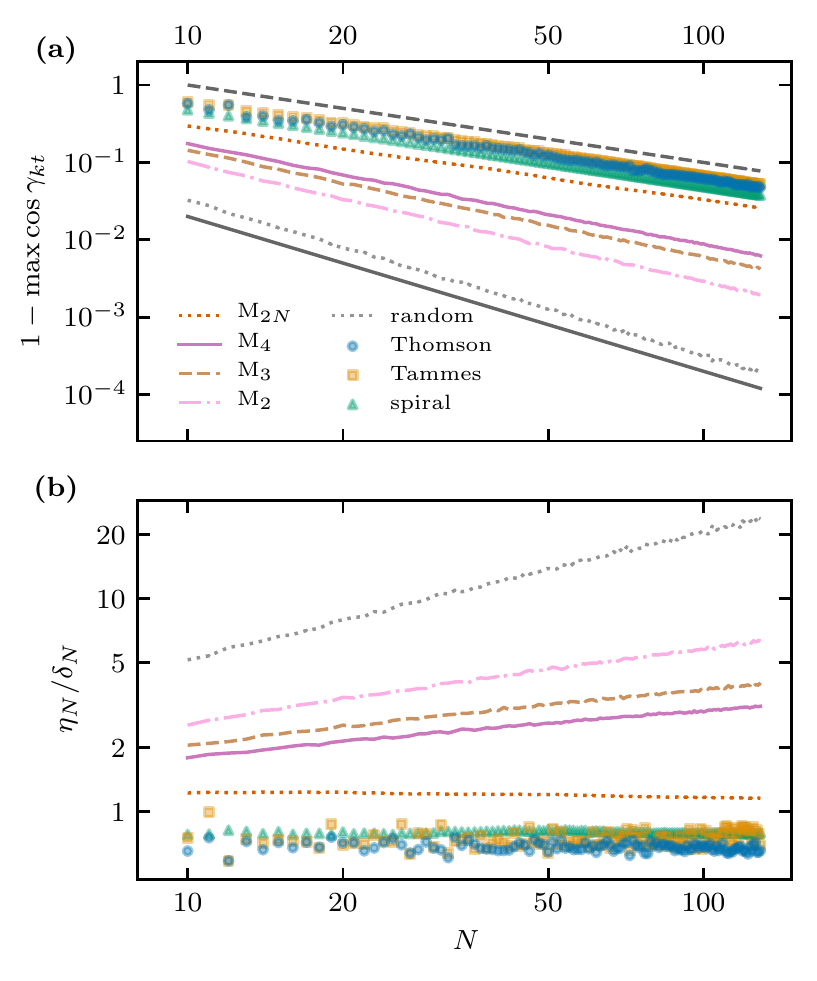}
\end{center}
\caption{{\bf (a)} Minimum distance between any two particles (in the form of $1-\max\cos\gamma_{kt}$) and {\bf (b)} mesh ratio $\eta_N/\delta_N$ in different types of distributions as a function of $N$. Dashed black line in panel (a) shows the scaling $N^{-1}$, obtained from Eq.~\eqref{eq:mindist} assuming a spherical hexagonal lattice, and the full black line shows the scaling $N^{-2}$, valid for uniform random distributions.}
\label{fig:5}
\end{figure}

There are also other measures often used for describing spherical distributions, concerned with their properties that are desirable for various applications. Most popular among these measures are uniformity (or equidistribution)~\cite{Kuipers}, quasi-uniformity~\cite{Hardin2016}, and the minimum energy~\cite{Brauchart2015}. While all the distributions considered in this work are uniform, it is instructive to take a look at their quasi-uniformity, which is defined as a bounded mesh ratio---ratio of covering radius $\eta_N$ and separation $\delta_N$. The latter is simply the minimum distance between any two particles in a distribution, $\delta_N=\min_{k,t\in N}\gamma_{kt}$, and is shown in Fig.~\ref{fig:5}a in the form of $1-\max\cos\gamma_{kt}$. We can see that the more ordered distributions possess a larger minimum distance between particles, which corresponds to a gap in their pair correlation function (e.g., Fig.~\ref{fig:1}b) and further translates into a vanishing structure factor at low $\ell$. Their power-law scaling goes as $N^{-1}$, as predicted by Eq.~\eqref{eq:mindist}, and this scaling gradually changes to $N^{-2}$ as distributions become more disordered. The increase in the negative exponent goes thus hand-in-hand with an increase in the hyperuniformity parameter $A_N$. This hints that at least for the tested scale-free distributions, the concept of hyperuniformity is connected to the minimum distance more than just qualitatively.

The other part of the mesh ratio of a distribution is the covering radius, which is defined as $\eta_N=\max_{\Omega\in\mathcal{S}^2}\min_{k\in N}|\Omega_k-\Omega|$~\cite{Hardin2016}. Divided by the minimum distance $\delta_N$, it gives the mesh ratio $\eta_N/\delta_N$, and the distributions for which the ratio remains bounded in the limit of $N\to\infty$ are termed quasi-uniform. Note that quasi-uniformity does not imply equidistribution, or vice versa. Figure~\ref{fig:5}b shows that, of the distributions considered in this work, the three ordered distribution (Thomson, Tammes, and spiral) and the M$_{2N}$ distribution can be considered quasi-uniform. The same four distributions are also the most hyperuniform, although the M$_{2N}$ distribution has a small but non-vanishing coefficient $A_N$, hinting again at the relatedness of these measures.

While one might be tempted to equate quasi-uniformity with hyperuniformity, there are several differences which make them stand apart. The first one is that, while quasi-uniformity can classify well different distributions when the mesh ratio is bounded~\cite{Hardin2016}, it does not provide much information when the ratio is unbounded. On the other hand, we can talk about the degrees of hyperuniformity as given by the parameters $A_N$ and $B_N$ also when the underlying distribution is not quasi-uniform. Moreover, the notion of quasi-uniformity breaks down when we have hierarchically ordered distributions which are locally clustered but ordered on the global scale (such as the ones formed by particles interacting with GEM-4 potential mentioned previously~\cite{Franzini2018}). In contrast, this is where our definition of hyperuniformity excels, as it is designed precisely to discover order at large length scales, no matter how the local structure of a distribution looks like.

\section{Conclusions}

The concept of hyperuniformity has since its introduction in 2003~\cite{Torquato2003} grown into an important tool to quantify order in terms of long-range correlations, bringing locally very different structures of crystals, quasicrystals, glasses, and jammed packings onto the same footing~\cite{Torquato2018}. Many recent publications---even within the last year---document an increased interest in the introduction of hyperuniformity for spherical distributions, and have made steps to adapt some part of the concept to their needs~\cite{Brauchart2018,Brauchart2018b,Meyra2018}, motivating us to develop a robust formalism. By utilizing the spherical structure factor, we have adapted the number variance, which differentiates hyperuniform and non-hyperuniform states through different asymptotic behaviour, to spherical geometry. While a lot of properties, such as the circumference-scaling of the number variance and the presence of a gap in the structure factor, have direct parallels with the Euclidean case, there are also crucial differences. Compactness of spherical domains puts a special importance to cases with small numbers of particles and to the asymptotic scaling of various measures with the number of particles. Geometry also causes a unique angular dependence of the cap number variance, which is universal, and quantifies hyperuniformity as the vanishing of the leading term in the asymptotic expansion.

Distinguishing between hyperuniform and non-hyperuniform systems on the surface of the sphere has numerous potential uses in addition to theoretical analysis of point distributions for purposes of Monte Carlo integration, data representation, and tessellation. It can be used to study spherical systems in the context of electrostatics of patchy surfaces, where both experiments and theory have revealed a long-range attraction between even overall neutral surfaces, locally charged in a mosaic-like structure of positively and negatively charged domains~\cite{Adar2017,ALB2013a}, and where hyperuniform and non-hyperuniform distributions of charge might behave differently. Hyperuniformity can also be applied to the study of biological structures, such as spherical viruses~\cite{Siber2012} or lipid rafts in vesicles~\cite{Jacobson2007}. It can aid in the classification of order and transitions in assemblies of colloidal shells~\cite{Giarritta1992,Giarritta1993,Fantoni2012,Subramaniam2005,Guerra2018,Hsu2005,Law2018} and can eventually be used to guide their design and properties~\cite{Dinsmore2002,Shah2009}. Moreover, the framework of hyperuniformity could help identify jammed states of matter on the sphere and jamming transitions~\cite{Goodrich2012,Goodrich2014,Ikeda2017}. In systems with an intrinsic length scale in their pair interaction potential~\cite{Franzini2018,Julija2011,Vest2014,Paquay2016,Vest2018}, clustering and ordering phase transitions can be investigated through the changes in the hyperuniformity parameters $A_N$ and $B_N$, as presented in this work. Naturally, at finite temperatures, thermal fluctuations also introduce a characteristic length scale that can drive phase transitions. In the future, generalizations of the framework of hyperuniformity on the sphere can be made to include two-phase heterogeneous media and continuous distributions; to generalize the approach to other geometries, such as periodic toroidal manifolds, negatively curved manifolds or even discrete realm of lattices and graphs; and to use the properties of the spherical structure factor to generate spherical distributions with desired properties, all following the examples from the Euclidean space~\cite{Torquato2018}.

\acknowledgments

We thank Julija Zavadlav for sharing with us the data on the spherical solutions of the hard-core--soft-shoulder potential and Stefano Franzini for sharing the data on the spherical solutions of the GEM-4 potential. We acknowledge financial support from the Slovenian Research Agency (research core funding Nos.\ P1-0055 (A.L.B.) and P1-0099 (S.\v C.), and research project No.\ J1-9149).

\appendix

\section{Analytical estimates}
\label{sec:appA}

For the more ordered particle distributions on the sphere, such as the solutions of the Thomson and Tammes problems, we can assume that the particles are arranged in a spherical hexagonal lattice~\cite{Saff1997}. Due to the obligatory $12$ pentagonal disclinations on the surface of a sphere, the average number of neighbours on such a lattice is $n=6-12/N$. Euler characteristic for the sphere demands that $N-E+F=2$, where we have $E=[12\times5+6(N-12)]/2$; we also see that $F=2N-4$. For a hexagonal lattice, we have $F$ equilateral triangles, and for large enough $N$ we can use the Euclidean formula to obtain $4\pi/F=(\min\gamma)^2\sqrt{3}/4$, where $\min\gamma$ is the minimum distance between two particles---the side of an equilateral triangle. It immediately follows that
\begin{equation}
\max\cos\gamma=\cos\left(\frac{16\pi}{\sqrt{3}(2N-4)}\right)\approx1-\frac{4\pi}{\sqrt{3}N},
\end{equation}
where we have moreover written $\cos\min\gamma=\max\cos\gamma$. This turns out to be a good fit for the position of the first peak of the pair distribution function, and is a known asymptotic result in the case of the solutions of the Tammes problem~\cite{Saff1997}.

\begin{figure}[!b]
\begin{center}
\includegraphics[width=\columnwidth]{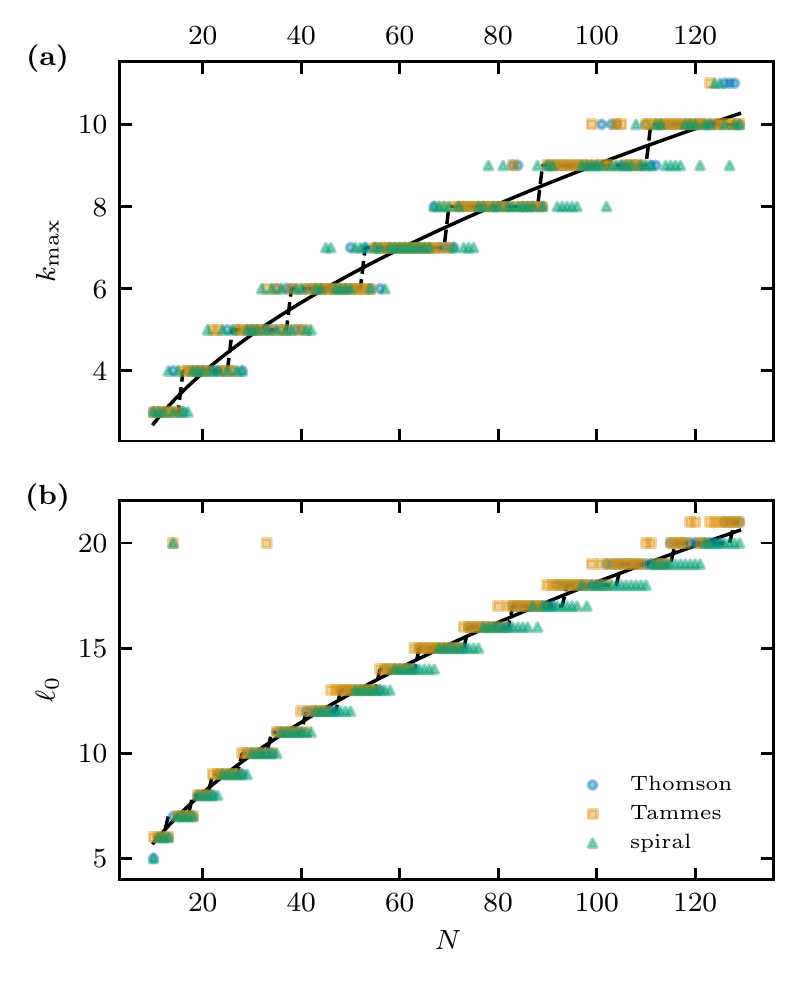}
\end{center}
\caption{ {\bf (a)} Number of peaks $k_{\max}$ in the angular dependence of the cap number variance $\sigma_N^2(\theta)$ of Thomson and Tammes particle distributions as well as of generalized spiral nodes. The peaks correspond to ``shells'' of neighbours and are indicative of an ordered lattice-like structure of the particles on the sphere. Full black line shows the approximation of Eq.~\eqref{eq:kmax}, obtained assuming a spherical hexagonal lattice; dashed black line shows its rounded (integer) values. The number of peaks in the cap number variance of different distributions is also approximate due to uncertainty of peak determination when $\theta\approx\pi/2$. {\bf (b)} The position $\ell_0$ of the first peak (maximum) in the spherical structure factor $S_N(\ell)$ of Thomson, Tammes, and spiral distributions. Full black line shows the approximation of Eq.~\eqref{eq:ell0}, which relates the position of the first peak in the spherical structure factor to the number of peaks in the angular dependence of the cap number variance, $\ell_0\approx2k_{\max}$, and dashed black line shows its rounded value.}
\label{fig:A1}
\end{figure}

If we instead approximate the spherical cap occupied by nearest neighbours of any given point as containing $n$ triangles, we get
\begin{equation}
n\frac{4\pi}{F}=2\pi(1-\cos\gamma),
\end{equation}
and so
\begin{equation}
\label{eq:mindist}
\max\cos\gamma=1-\frac{2n}{F}=1-\frac{2(6-12/N)}{2N-4}=1-\frac{6}{N}.
\end{equation}
This dependence fits very well to the minimum distance between particles given by the solutions of the Thomson and Tammes problems, where the numerical factor in the dependence is $\sim5.5$ instead of $6$ (scaling factor in Fig.~\ref{fig:5}a). In deriving the equation, the topological restriction for the average number of neighbours was taken into account. We can extend this fit to the $k$-th layer of particles, obtaining
\begin{equation}
\cos\gamma_k=\cos(k\min\gamma)=\cos\left(k\arccos\left(1-\frac{6}{N}\right)\right),
\end{equation}
This expression also allows us to estimate the number of peaks in the angular dependence of the cap number variance for ordered distributions as
\begin{equation}
\label{eq:kmax}k_{\max}=\frac{\pi}{\arccos(1-6/N)}
\end{equation}
which can be expanded for large $N$ to yield $k_{\max}\approx\pi\sqrt{N/12}$. Equation~\eqref{eq:kmax} provides an almost perfect estimate for the number of shells that form around particles in the Thomson and Tammes distributions, as is demonstrated in Fig.~\ref{fig:A1}a.

By considering a structure factor with a single non-zero element at $\ell=\ell_0$, $S_N(\ell_0)=1$, we can obtain the corresponding cap number variance from Eq.~\eqref{eq:var} and show that its number of peaks scales as $k_{\max}\approx\ell_0/2$. Thus, given a prominent first peak in a structure factor of an ordered distribution of $N$ particles, we can reasonably approximate its position for large $N$ as
\begin{equation}
\label{eq:ell0}\ell_0=\frac{\pi}{\sqrt{3}}\sqrt{N}
\end{equation}
This relation is shown in Fig.~\ref{fig:A1}b, and turns out to be a very good approximation for both Thomson and Tammes distributions. Generalized spiral nodes, on the other hand, retain the functional dependence of $\ell_0\propto\sqrt{N}$, but the peak occurs slightly before the one predicted by Eq.~\eqref{eq:ell0}, meaning that the prefactor of the dependence is somewhat smaller than predicted.

Note that the approximation for the minimum distance between particles derived in this Appendix is similar to the approximations based on the disco ball model~\cite{Hardin2016} and zonal equal area partitioning~\cite{Leopardi2006}, where, however, the underlying lattice is assumed to be square. For instance, the estimate from Ref.~\cite{Leopardi2006} based on zonal equal area partitioning yields a large $N$ approximation for minimum distance as $\max\cos\gamma\approx1-2\pi/N$. The factor given by this approximation proves to be a slightly worse fit for ordered distributions than the one derived from a hexagonal lattice.

\section{Derivation of cap number variance in terms of spherical structure factor}
\label{sec:appB}

To derive Eq.~\eqref{eq:var} that connects the spherical structure factor with the cap number variance, we begin by taking the expressions for the density distribution of $N$ particles, $\rho(\Omega)$, and the spherical cap centred at $\Omega_C$ with an opening angle $\theta$, $C(\Omega_C,\theta)$, and we expand both functions in terms of spherical harmonics. We denote the thusly obtained multipole expansion coefficients by $\rho_{lm}$ and $c_{lm}$, respectively. This allows us to write the first term in Eq.~\eqref{eq:var1} as
\begin{eqnarray}
\label{eq:n2}
\left\langle\vert N(\theta)\vert^2\right\rangle
\nonumber&=&\left\langle\sum_{l,m}\rho_{lm}c^*_{lm}\sum_{l',m'}\rho^*_{l'm'}c_{l'm'}\right\rangle\\
&=&\sum_{l,m}\sum_{l',m'}\rho_{lm}\rho^*_{l'm'}\left\langle c_{l'm'}c^*_{lm}\right\rangle,
\end{eqnarray}
since the average is taken over all possible orientations (centres) of the spherical cap and thus affects solely the spherical cap coefficients; in the derivation we have also used the orthogonality of spherical harmonics. We next use the Wigner D-matrices $D_{mm'}^{(l)}(\bm{\omega})$ to represent spherical caps with different centres $\Omega_C$ through rotations of the coefficients of a single spherical cap centred at the north pole, $\Omega_C={\bf 0}$~\cite{ALB2013a}. Thus we can write
\begin{eqnarray}
\left\langle c_{l'm'}c^*_{lm}\right\rangle
\nonumber&=&\left\langle\sum_{k,k'}D_{km}^{(l)*}(\bm{\omega})\tilde{c}^*_{lk}D_{k'm'}^{(l')}(\bm{\omega})\tilde{c}_{l'k'}\right\rangle\\
&=&\sum_{k,k'}\tilde{c}^*_{lk}\tilde{c}_{l'k'}\left\langle D_{km}^{(l)*}(\bm{\omega})D_{k'm'}^{(l')}(\bm{\omega})\right\rangle
\end{eqnarray}
where $\tilde{c}_{lm}$ are the expansion coefficients of the spherical cap centred at the north pole and $\bm{\omega}=(\alpha,\beta,\gamma)$ are the Euler angles required to rotate the pole-centred cap to (some) $\Omega_C$~\cite{ALB2013a}. We can then express the average over all spherical cap centres as the average over all rotations $\bm{\omega}$ of the pole-centred cap, and we obtain the known result
\begin{eqnarray}
\left\langle D_{km}^{(l)*}(\bm{\omega})D_{k'm'}^{(l')}(\bm{\omega})\right\rangle
\nonumber&=&\frac{1}{8\pi^2}\int\mathrm{d}\bm{\omega}\, D_{km}^{(l)*}(\bm{\omega})D_{k'm'}^{(l')}(\bm{\omega})\\
&=&\frac{\delta_{ll'}\delta_{mm'}\delta_{kk'}}{2l+1}.
\end{eqnarray}
We can now insert the above expressions into Eq.~\eqref{eq:n2}, from where it follows
\begin{eqnarray}
\left\langle\vert N\vert^2\right\rangle
\nonumber&=&\sum_{l,m}\sum_{l',m'}\rho_{lm}\rho^*_{l'm'}\frac{\delta_{ll'}\delta_{mm'}}{2l+1}\sum_k\vert \tilde{c}_{lk}\vert^2\\
&=&\sum_l\frac{1}{2l+1}\sum_m\vert\rho_{lm}\vert^2\sum_k\vert \tilde{c}_{lk}\vert^2.
\end{eqnarray}
Taking into account also the expression for multipole magnitudes [Eq.~\eqref{eq:Sl}], and denoting by $Q_l$ and $C_l$ the multipole magnitudes of the particle distribution and spherical cap, respectively, we obtain
\begin{equation}
\left\langle\vert N\vert^2\right\rangle=\frac{1}{(4\pi)^2}\sum_l(2l+1)\,Q_l^2\,C_l^2.
\end{equation}

We can use a similar derivation to show that the second term in Eq.~\eqref{eq:var1}, $\vert\langle N\rangle\vert^2$, is identical to $0$ for all $l\geqslant1$ due to the averaging of a single Wigner D-matrix over all possible rotations. What is more, the monopole ($l=0$) terms of $\left\langle\vert N\vert^2\right\rangle$ and $\vert\langle N\rangle\vert^2$ cancel out, and we thus obtain
\begin{equation}
\label{eq:var2}
\sigma_N^2(\theta)=\frac{N}{(4\pi)^2}\sum_{l\geqslant1}(2l+1)\,S_N(l)\,C_l^2,
\end{equation}
where we have also used the definition of the spherical structure factor $S_N(l)$ [Eq.~\eqref{eq:sl}]. This equation is similar to Eq.~(S-24) in Ref.~\cite{Pilleboue2015} where a similar derivation has been made for different purposes; note the different factor due to different definitions of spherical harmonics used. It is also easy to further obtain the multipole expansion coefficients of a spherical cap with opening angle $\theta$ and centred at the north pole~\cite{Pilleboue2015,ALB2011}
\begin{equation}
\tilde{c}_{lm}=\sqrt{\frac{\pi}{2l+1}}\left[P_{l+1}(\cos\theta)-P_{l-1}(\cos\theta)\right]\,\delta_{m0},
\end{equation}
giving
\begin{equation}
C_l^2=\frac{4\pi^2}{(2l+1)^2}\left[P_{l+1}(\cos\theta)-P_{l-1}(\cos\theta)\right]^2.
\end{equation}
Inserting this into Eq.~\eqref{eq:var2}, we finally obtain Eq.~\eqref{eq:var}. We have thus been able to express the cap variance of a spherical distribution solely in terms of its structure factor.

\section{Asymptotic expansion of cap number variance}
\label{sec:appC}

In order to derive Eqs.~\eqref{eq:var_sin} and \eqref{eq:expansion}---the functional dependence of the cap number variance when the structure factor has a gap at low multipole numbers---we will utilize the asymptotic behaviour of Legendre polynomials as $l\to\infty$,
\begin{equation}
P_l(\cos\theta)\asymp\frac{2}{\sqrt{2\pi l\sin\vartheta}}\cos\left[\left(l+\frac{1}{2}\right)\theta-\frac{\pi}{4}\right].
\end{equation}
The contribution of a single multipole number $\ell_0$ to the sum in Eq.~\eqref{eq:var} will serve as a Green's function to construct a general structure factor and study its role in the cap number variance:
\begin{eqnarray}
\label{eq:green}
\nonumber G(\ell_0)&=&\frac{(P_{\ell_0+1}-P_{\ell_0-1})^2}{2\ell_0+1}\\
&\asymp&\frac{2}{\pi \ell_0^2}\sin\theta\left[1-\sin((2\ell_0+1))\theta\right],
\end{eqnarray}
where we have approximated $\ell_0\pm1\approx\ell_0$ in the denominator. These approximations rely on the limit of large $\ell$---the relative error of the approximation scales as $1/\ell$. The first term in the square brackets of Eq.~\eqref{eq:green} scales with the normalized circumference of the cap, $\sin\theta$, and the second term is the oscillatory contribution of a single peak in the structure factor. 

Crystalline distributions, such as the Tammes and Thomson distributions, can be approximately described as a series of periodic sharp peaks spaced by $\ell_0$, with peak amplitudes equal to $\ell_0$ to ensure that the average asymptotic structure factor tends to unity. The spherical cap variance for such a peaked distribution reduces to the Basel problem:
\begin{eqnarray}
\label{eq:sigma_peaked}
\nonumber\sigma^2_N(\theta)&=&\frac{N}{4}\sum_{k=1}^\infty \ell_0\,G(k\ell_0)\\
&=&\frac{N}{4}\frac{2}{\pi\ell_0}\sin\theta\times\frac{\pi^2}{6}+\xi(\ell_0,\theta).
\end{eqnarray}
The residual term $\xi(\ell_0,\theta)$ captures a Fourier sum over the oscillatory contributions, with frequency determined by the first peak $\ell_0$ and higher harmonics only affecting the precise shape of the waveform. Excepting distributions with perfect symmetry, peaks in the structure factor are not sharp and the oscillatory residual is suppressed. Exact peak width depends on the details of the distribution and varies with $N$, as more symmetric distributions (icosahedral, tetrahedral, and octahedral~\cite{Lorman2007,ALB2013a}) tend to have sharper peaks.

\begin{figure}[!b]
\begin{center}
\includegraphics[width=\columnwidth]{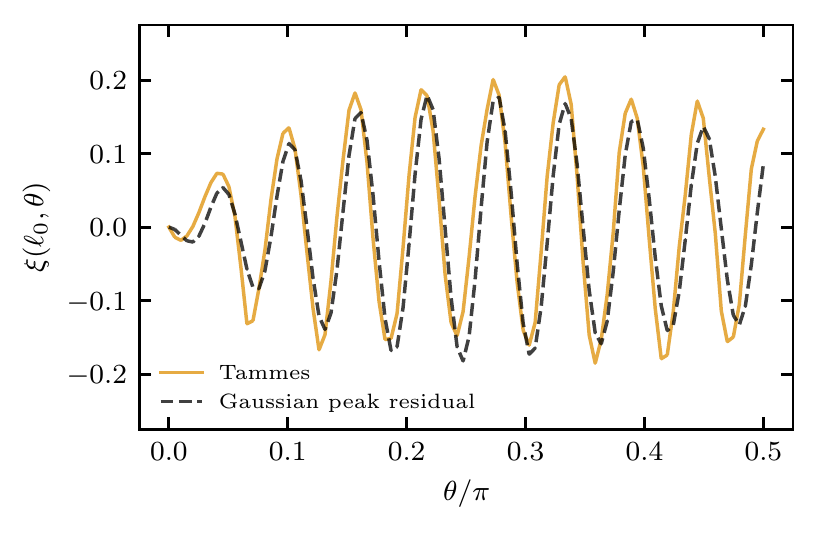}
\end{center}
\caption{Residual cap number variance $\xi(\ell_0,\theta)$ (full line) of a Tammes distribution with $N=85$ particles, obtained by subtracting the fitted cap number variance $\sigma_N^2(A,B;\theta)$ [Eq.~\eqref{eq:var_fit}] from the actual cap number variance $\sigma_N^2(\theta)$. Dashed line shows a fit of Eq.~\eqref{eq:residue} to the residual term, and includes the contribution of a single peak with a Gaussian profile in the spherical structure factor at $\ell_0=\pi\sqrt{N}/\sqrt{3}$ to the cap number variance. The fit parameters of the residual term [Eq.~\eqref{eq:residue}] are $C=0.49$ for the prefactor and $\lambda=1.39$ for the exponent; the parameters $A$ and $B$ of the subtracted term are the same as in Fig.~\ref{fig:4}.}
\label{fig:C1}
\end{figure}

For a single peak in the structure factor which has a Gaussian profile proportional to $\exp(-\lambda(\ell-\ell_0)^2)$, the normalization to intensity $\ell_0$ that we have assumed in the derivation of Eq.~\ref{eq:sigma_peaked} and the summation over $\ell$ together yield a multiplicative envelope expressed with Jacobi elliptic theta function $\vartheta$:
\begin{equation}
\label{eq:residue} 
\xi(\ell_0,\theta)\approx -C\,\frac{N}{2\pi \ell_0} \frac{\vartheta(\theta;e^{-\lambda})}{\vartheta(0;e^{-\lambda})}\sin\theta \sin((2\ell_0+1)\theta).
\end{equation}
The envelope suppresses oscillations starting at $\theta=\pi/2$, and the suppression progresses outwards to smaller angles as $\lambda$ is increased. The multiplicative factor $C$ in Eq.~\eqref{eq:residue} is an additional fitting parameter that compensates for the inexact nature of this approximation, mostly due to unequal distribution of intensity between first and other peaks. The result qualitatively matches the oscillatory behaviour observed in Fig.~\ref{fig:4} for Thomson, Tammes, spiral, and M$_{2N}$ distributions. Figure~\ref{fig:C1} shows a comparison between the residual oscillations of the cap number variance of a Tammes distribution with $N=85$ particles and its fit to Eq.~\eqref{eq:residue}. The fit describes the behaviour of the residual oscillations extremely well, but at the same time we can observe its sensitivity to the choice of $\ell_0$---which was chosen according to the approximation in Eq.~\eqref{eq:ell0} and is, as Fig.~\ref{fig:A1}b shows, not always exact.

If the spherical structure factor can be approximated with a continuous function, summation in the expression for the cap number variance can be replaced by an integral of the Green's function, Eq.~\eqref{eq:green}. The approximation is invalid for low $\ell$, as witnessed by the fact that for $S(\ell)=1$, the sum reduces to $\sigma_N^2=N\sin^2\theta/4$; consequently, the low-frequency contribution has to be treated separately:
\begin{equation}
\label{eq:integral}
\sigma^2_N=\frac{N}{4}\left\{\overline{S}_N(0)\sin^2\theta+\int_0^\infty \left[S_N(\ell)-\overline{S}_N(0)\right]G(\ell)\,{\rm d}\ell\right\},
\end{equation}
ignoring the oscillatory residual. As an example, consider a spherical structure factor in the form of a commonly used combination of a gap (Heaviside $H$ function) and a Dirac delta peak:
\begin{equation}
\label{eq:sf_gd}
S(\ell)=H(\ell-\ell_0)+\frac{\ell_0}{2}\,\delta(\ell-\ell_0),
\end{equation}
where the peak normalization compensates for the gap, preserving normalization of the structure factor [see Eq.~\eqref{eq:parseval}]. The first contribution can be calculated using the integral in Eq.~\eqref{eq:integral} and the second is already given by the dominant term of the Green's function [Eq.~\eqref{eq:green}], yielding the result
\begin{equation}
\sigma^2_N=\frac{N}{4}\frac{3}{\pi \ell_0}\sin\theta+\xi(\theta).
\end{equation}
The numerical prefactor of this result is slightly smaller than the result obtained for a series of periodic peaks [Eq.~\eqref{eq:sigma_peaked}], indicating that the fitting parameter $B$ can differentiate between different types of hyperuniform order. Note also that the prefactor in Eq.~\eqref{eq:ell0} for the dependence of the peak position $\ell_0$ on $N$, derived in Appendix~\ref{sec:appA}, relies on the notion of nearest neighbours in equally spaced crystalline-like distributions, and may be different in the case of a distribution with the structure factor in the form of a gap combined with a delta peak [Eq.~\eqref{eq:sf_gd}], just as the dependence of the minimum distance on $N$ is different for different distributions (see Fig~\ref{fig:5}a).

The common scaling with $N$ and $\theta$ allows us to consider the prefactors to $\sin\theta$ and $\sin^2\theta$ as empirical parameters that quantify the degree of order in the distribution, and in this way we arrive at the final asymptotic form in Eq.~\eqref{eq:var_fit}. We take the model of the most ordered distribution [Eq.~\eqref{eq:var_gapped}] combined with hexagonal lattice approximation [Eq.~\eqref{eq:ell0}] as the reference against which all distributions are compared, normalizing the scale of the parameter $B$.

\section{Analysed spherical distributions}
\label{sec:appD}

In this work, we use several different distribution types to study the relation between their spherical structure factor and cap number variance, and use this analysis to derive a classification of hyperuniformity on the sphere.

As examples of ordered distributions, which include minimum energy configurations resulting from power-law and logarithmic interaction potentials between particles~\cite{Saff1997,Hardin2016}, we study the solutions of the Thomson and Tammes problems. These represent two extremes of interaction potentials: in the former case, particles interact with each other through long-range electrostatic repulsion; in the latter case, particles interact solely through a close-range hard-core repulsion. We have obtained the solutions of the Thomson problem from the Cambridge Cluster Database~\cite{CCD,Wales2006} and the solutions of the Tammes problem from N.\ Sloane's Spherical codes repository~\cite{SloaneWEBa}. Solutions of both problems exhibit a range of different symmetries, from highly symmetric icosahedral and octahedral configurations to configurations with a single $n$-fold rotation symmetry; some lack any kind of symmetry whatsoever. In addition to Thomson and Tammes distributions, we consider generalized spiral nodes, where a generating spherical spiral is used to define a sequence of points on the sphere. Such distributions are often used in numerical analysis and approximation theory as they possess several desirable properties, such as quasi-uniformity and equidistribution, and they approach equal-area distributions in the limit of large $N$~\cite{Hardin2016}.

We also consider several types of randomly-generated distributions. The basic form is, of course, a completely disordered, uniform random distribution, whose form is amenable to an analytical derivation of both the spherical structure factor as well as the cap number variance. In addition, we generate distributions based on Mitchell's best candidate algorithm~\cite{Mitchell1991,ALB2018b}, which approximates Poisson disc sampling and blue noise on the sphere. Mitchell's algorithm uses resampling to place the particles on the sphere: the $i+1$-th particle is placed randomly on the sphere in $K$ trials, and the position which is the furthest away from all $i$ previously placed particles is retained. This procedure is repeated until all $N$ particles have been placed. We will consider the cases where we allow each particle a small, fixed number of trials ($K=2$, $3$, and $4$), as well as cases where the number of trials depends on $N$, $K=2N$. We label the different Mitchell-type distributions as $M_K$.

To summarize, in this work we use several different scale-free distributions with varying ranges of $N$, where the latter depend either on the availability of the data or on their ease of generation:
\begin{itemize}
\item Thomson distribution, $N\in[10,580]$;
\item Tammes distribution, $N\in[10,130]$;
\item generalized spiral distribution, $N\in[10,500]$;
\item M$_K$ ($K=2$, $3$, $4$, $2N$) distributions, $N\in[10,150]$;
\item uniform random distribution, $N\in[10,150]$;
\end{itemize}
Due to the nature of construction of randomly-generated distributions, we generate $1000$ different samples for each $N$ and use these to obtain ensemble averages of the quantities of interest (such as spherical structure factor and cap number variance). While we have used the full range of available $N$ for each distribution to obtain the results presented in this work, we have also checked that our conclusions still hold if we limit our entire dataset to $N\in[10,130]$, the smallest range of available data (Tammes distribution).

Lastly, we mention the use of data provided to us by Julija Zavadlav~\cite{Julija2011} (HC-SS potential) and Stefano Franzini~\cite{Franzini2018} (GEM-4 potential), which encompass distributions of particles interacting with soft potentials possessing an internal length scale. Such distributions require a more detailed analysis as both the variation in $N$ and $\rho$ must be examined systematically, which is beyond the scope of our paper.

\bibliography{references}

\end{document}